\documentclass[preprintnumbers,amsmath,amssymb,aip,twocolumn,showpacs]{revtex4}

\usepackage{graphicx}
\usepackage{dcolumn}
\usepackage{bm}
\usepackage{mdwlist}
\usepackage{setspace}
\usepackage{natbib}
\usepackage{amssymb}
\usepackage{amsmath}
\usepackage{paralist}
\usepackage{enumitem}
\usepackage{fouridx}
\usepackage{nicefrac}
\DeclareMathOperator{\erf}{erf}
\DeclareMathOperator{\sn}{sn}
\DeclareMathOperator{\K}{K}
\DeclareMathOperator{\csch}{csch}
\DeclareMathOperator{\arcsinh}{arcsinh}
\DeclareMathOperator{\Ein}{Ein}
\DeclareMathOperator{\im}{Im}
\DeclareMathOperator{\F}{F}
\DeclareMathOperator{\Ei}{Ei}
\DeclareMathOperator{\Ry}{Ry}

\begin{document}

\title{Heating mechanisms in radio frequency driven ultracold plasmas}

\author{P. W. Smorenburg}
\author{L. P. J. Kamp}
\author{O. J. Luiten}
\email{o.j.luiten@tue.nl}
\affiliation{Coherence and Quantum Technology (CQT), Eindhoven University of Technology, PO Box 513, 5600 MB Eindhoven, The Netherlands}

\date{\today}

\begin{abstract}
Several mechanisms by which an external electromagnetic field influences the temperature of a plasma are studied analytically and specialized to the system of an ultracold plasma (UCP) driven by a uniform radio frequency (RF) field. Heating through collisional absorption is reviewed and applied to UCPs. Furthermore, it is shown that the RF field modifies the three body recombination process by ionizing electrons from intermediate high-lying Rydberg states and upshifting the continuum threshold, resulting in a suppression of three body recombination. Heating through collisionless absorption associated with the finite plasma size is calculated in detail, revealing a temperature threshold below which collisionless absorption is ineffective.
\end{abstract}


\maketitle

\section{Introduction}\label{sec0}
Conventional plasmas are formed when atoms are ionized by strong electric fields or collisions with other particles. Due to the large excess energy inherent in such ionization processes, the resulting electron temperature is typically comparable to the ionization potential, which is on the order of an electronvolt, equivalent to some $10^4$ K. In marked contrast, ultracold neutral plasmas (UCPs), created by photo-ionization of a cloud of laser-cooled atoms \cite{Killian1}, have an electron temperature close to $1$ K. UCPs typically consist of some $10^8$ singly-ionized atoms localized in a millimeter-sized cloud of Gaussian density profile, with a correspondingly low particle density \cite{Killian2}. The combination of low temperature and low density makes UCPs unique plasma systems. They can be close to the strongly-coupled regime where the Coulomb interaction energy between the particles exceeds the thermal energy, as is quantified by the coupling parameter
\begin{align}
\Gamma=\frac{e^2}{4\pi\epsilon_0r_{w}k_BT}\label{0.1}
\end{align}
exceeding unity, where $e$ is the electron charge, $\epsilon_0$ the vacuum permittivity, $k_B$ Boltzmann's constant, $T$ the plasma temperature, and $r_{w}=\left[3/(4\pi n)\right]^{1/3}$ the Wigner-Seitz radius with $n$ the number density. Due to their high coupling parameter, UCPs behave in many respects similar to strongly-coupled plasmas near solid state density, such as laser-ionized atomic clusters \cite{Fennel} or thin films \cite{Hatchett}, inertial confinement fusion targets \cite{Hu} and astrophysical plasmas \cite{Chabrier}. The dynamics of solid state density plasmas, however, takes place at the time scale of the inverse plasma frequency, which lies in the attosecond to femtosecond regime. This seriously complicates diagnostics. In contrast, UCPs evolve on the time scale of picoseconds to microseconds. This enables excellent time-resolved diagnostic techniques, including charged particle detection \cite{Li}, absorption imaging \cite{Simien} and fluorescence monitoring \cite{Cummings}. In addition, the careful preparation and ionization of atomic clouds allows accurate control over the initial temperature, density profile, and ionization state. UCPs may therefore serve as versatile and experimentally accessible model systems for high-density plasmas that are difficult to diagnose.\\

An important class of experiments on solid state density plasmas involves plasmas created by laser irradiation of atomic clusters in a gas jet. Characteristic of these experiments is that the laser pulse length is comparable to the lifetime of the plasma. Therefore the studied system typically consists of a cluster plasma that is not only near to strongly-coupled, but is also strongly driven by a radiation field. This leads to complicated dynamics that is difficult to unravel \cite{Fennel}. Research on laser-cluster interaction would therefore benefit from UCP experiments in which this interaction is mimicked. Since atomic clusters are typically smaller than the laser wavelength, the appropriate model system is an UCP driven by a strong radio-frequency (RF) field. Interpretation of observations in such experiments on RF driven UCPs, however, requires a detailed understanding of the mechanisms by which the RF field and the UCP interact. In this paper, we consider how the RF field influences the plasma temperature, both directly through RF energy absorption mechanisms and indirectly through modification of the three body recombination process, the latter being a main heat source in UCPs.\\

In current UCP experiments, RF fields are used in a diagnostic way to probe plasma modes. Plasma resonance can be detected as an increased yield of electrons leaving the UCP \cite{Kulin}. Combined with knowledge of the mode properties \cite{Bergeson}, this can be used to determine the plasma density and expansion as a function of time. Using the same technique, the presence of acoustic or Tonks-Dattner modes in an UCP has been observed in addition to the fundamental mode \cite{Fletcher}. In these experiments, the collective response of the plasma electrons to the RF field has been studied in quite some detail. However, the RF amplitude is kept low to avoid disturbances other than plasma resonances, and little attention is paid to other interaction mechanisms. Nevertheless, as we will describe in this paper, the RF field influences the plasma also via incoherent processes. In their Tonks-Dattner modes experiment, Fletcher et al. \cite{Fletcher} indeed observe the onset of field-induced effects at large probing amplitudes. Although lower RF amplitudes justify the use of standard plasma quantities, such as the Spitzer collision frequency applied in the interpretation of the fundamental plasma resonance measurements \cite{Bergeson}, or the Debye length mentioned in support of the analysis of the Tonks-Dattner modes \cite{Fletcher}, one should be aware of the possible high-amplitude modifications of such quantities induced by the RF field. Finally, the expansion of an UCP is driven by the thermal pressure of the electrons. It is therefore important to understand the various ways in which the RF field contributes to the heat budget of the plasma.\\

In this paper, we take the electric field strength $E_0$ in the plasma as a given quantity, and consider what influence this field has on several microscopic processes. For underdense plasmas, $E_0$ is approximately equal to the externally applied RF field. For denser plasmas, $E_0$ may be significantly enhanced by the polarization field generated by the plasma itself. This is particularly relevant under conditions of resonance with plasma modes, in which case the absorption of RF energy by the UCP is dominated by the strong dependence of $E_0$ on the driving frequency \cite{Twedt}. The determination of the frequency response of the UCP, and hence the polarization fields, is actively being studied \cite{Bergeson,Fletcher,Twedt,Lyubonko}, but is outside the scope of this paper. Nevertheless, our results may be directly applied once $E_0$ is known.\\

This paper is organized as follows. We consider two mechanisms by which the UCP can directly absorb energy from the RF field: collisional absorption and collisionless absorption due to the finite size of the plasma. The first of these has been studied extensively already in other contexts \cite{Silin,Catto,Dawson,Decker,MulserBall,Osborn,Brysk,Bornath,Kull,Pert,Shima}. In Section \ref{sec1}, we therefore only cite the main results from literature and discuss their relevance for RF driven UCPs. In Section \ref{sec2}, we study the process of three body recombination in the presence of an RF field, and show that the recombination rate can be strongly suppressed by the field. Next, in Section \ref{sec3}, we consider the collisionless absorption mechanism mentioned above, which has been mainly studied in the context of solid-state density plasmas \cite{Blocki,Yannouleas,Brunel,Megi,Zaretsky,Kostyukov,MulserOsc,Kundu}. We show that the approximations usually adopted are not appropriate for UCPs. We provide an improved description by specializing a derivation of the collisionless absorption rate due to Zaretsky et al. \cite{Zaretsky} to the case of UCPs. We conclude and summarize in Section \ref{sec4}.

\section{Collisional absorption}\label{sec1}
\subsection{Collision frequency}\label{sec1.1}
At low to moderate RF field strengths, the energy absorption of a plasma is dominated by collisional absorption, or inverse Bremsstrahlung \cite{MulserBoek}. The physical cause of the absorption is that individual electrons, oscillating due to the RF field, deflect in the Coulomb fields of the approximately stationary ions, resulting in a net energy gain. The average effect of the Coulomb fields can be described phenomenologically as an effective frictional force $\bm{F}=-m\nu_\text{ei}\bm{v}$ in the equation of motion of the electron, and the energy absorption rate per electron by the power $P_\text{ei}=-\langle\bm{F}\cdot\bm{v}\rangle$. Here, $m$ is the electron mass, $\nu_\text{ei}$ is the effective electron-ion collision frequency, and $\bm{v}$ is the electron velocity. Expressing the velocity in terms of the driving electric field gives \cite{Silin}
\begin{align}
P_\text{ei}=2\nu_\text{ei}U_p,\label{1.1}
\end{align}
where $U_p=(eE_0)^2/(4m\omega^2)$ is the quiver energy, or ponderomotive potential, in the RF field with amplitude $E_0$ and frequency $\omega$. Here and in the remainder, we assume a linearly polarized RF field, and absorb any field enhancement due to plasma resonance in the magnitude $E_0$. Importantly, Eq. (\ref{1.1}) defines the collision frequency as merely a scaled absorption rate, rather than predicting the absorption from a predetermined collision frequency. Consequently, $\nu_\text{ei}$ is not necessarily equal to the Spitzer collision frequency \cite{Spitzer}
\begin{align}
\nu_S=\sqrt{\frac{2}{3\pi}}\omega_p\Gamma^{3/2}\ln\Lambda,\label{1.2}
\end{align}
which is commonly used for plasmas without RF fields. Nevertheless, the collision frequency Eq. (\ref{1.2}) is sometimes used for driven plasmas as well, and also in the context of RF absorption by UCPs \cite{Bergeson,Ronghua}. In Eq. (\ref{1.2}), singly ionized atoms are assumed, $\omega_p$ is the plasma frequency, and $\ln\Lambda$ is the Coulomb logarithm that will be discussed below. \\

Underlying any calculation of the collisional absorption rate is some model for the scattering of an electron by the Coulomb field of an ion, which generally depends on the electron velocity. Because two velocity scales are involved, namely the thermal velocity $v_\text{th}=\sqrt{k_BT_e/m}$ and the quiver velocity magnitude $v_\text{osc}=eE_0/(m\omega)$, the collision frequency depends on the ratio $v_\text{osc}/v_\text{th}$. Here, $T_e$ is the electron temperature of the plasma. The effective collision frequency has been calculated first by classical kinetic theory using the Landau collision integral \cite{Silin,Catto}. The result can be written as \cite{Brysk}
\begin{align}
\nu_\text{ei}=\nu_S\cdot\fourIdx{}{2}{}{2}{\F}\!\!\left(\frac{3}{2},\frac{3}{2};2,\frac{5}{2};-\frac{v_\text{osc}^2}{2v_\text{th}^2}\right),\label{1.3}
\end{align}
where $\fourIdx{}{2}{}{2}{\F}$ denotes the generalized hypergeometric function \cite{Andrews} that has the limiting forms
\begin{align}
\fourIdx{}{2}{}{2}{\F}\!\left(\dots\right)\approx\!\left\{
\begin{array}{ll}
1 &v_\text{osc}\ll v_\text{th}\\
6\sqrt{\frac{2}{\pi}}\left(\frac{v_\text{th}}{v_\text{osc}}\right)^3\left[\ln\left(\frac{v_\text{osc}}{2v_\text{th}}\right)+1.0\right] &v_\text{osc}\gg v_\text{th}.\label{1.4}
\end{array}\right.
\end{align}
More advanced and alternative calculations largely confirm these results \cite{Dawson,Decker,MulserBall,Osborn,Brysk,Bornath}.\\

\begin{figure}
\includegraphics[width=\columnwidth]{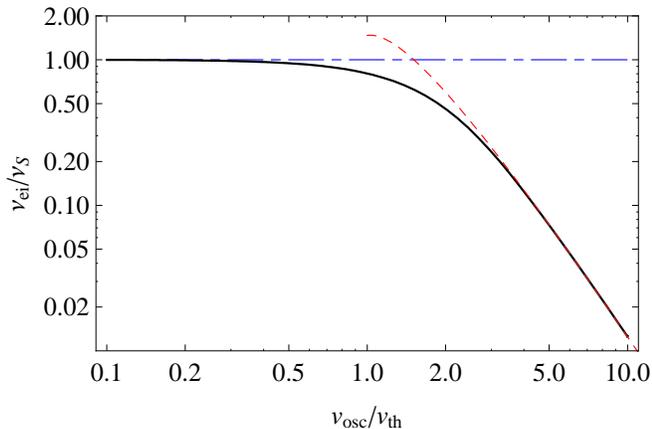}
\caption{\label{fig-2} (Color online) Effective electron-ion collision frequency for collisional absorption scaled to the Spitzer collision frequency, as a function of the ratio of quiver velocity to thermal velocity. Solid black line: collision frequency Eq. (\ref{1.3}); dash-dotted blue line: weak field limit $v_\text{osc}\ll v_\text{th}$ given by Eq. (\ref{1.4}); dashed red line: strong field limit $v_\text{osc}\gg v_\text{th}$ given by Eq. (\ref{1.4}).}
\end{figure}

The collision frequency of Eq. (\ref{1.3}) is plotted in Fig. \ref{fig-2} as a function of the velocity ratio. In RF experiments with UCPs, this ratio can vary over the full range $v_\text{osc}\ll v_\text{th}$ to $v_\text{osc}\gg v_\text{th}$ \cite{Kulin}. The decrease of the collision frequency for increasing $v_\text{osc}$ can be understood physically from the well-known fact that the Rutherford scattering cross section for an electron by an ion is inversely proportional to the fourth power of the relative velocity, so that driving the plasma stronger makes the electrons less susceptible to deflections and hence to energy gain. Note that the Spitzer frequency Eq. (\ref{1.2}) with Eq. (\ref{0.1}) substituted is proportional to $v_\text{th}^{-3}$, while the second line of Eq. (\ref{1.4}) contains the factor $\left(v_\text{th}/v_\text{osc}\right)^3$. Effectively, therefore, and apart from a logarithmic factor, the content of Eq. (\ref{1.3}) is that the thermal velocity is replaced by the quiver velocity in the collision frequency when $v_\text{osc}\gg v_\text{th}$. In fact, this effect is such that the collision absorption rate $P_\text{ei}$ given by Eq. (\ref{1.1}) decreases with field strength as $E_0^{-1}$ rather than increases, which is a well-known phenomenon in laser-plasma physics \cite{MulserBoek}. This behavior is not only relevant in situations where large RF field strengths are applied, but also when UCPs are driven resonantly. This is because the electric field $E_0$ is strongly enhanced at densities for which the plasma frequency equals the RF frequency. In particular, the amplitude of the electron oscillations is then limited by the dominant damping mechanism, which in view of Fig. \ref{fig-2} may no longer be collisional absorption. For sufficiently small $\nu_\text{ei}$, excitation of plasma waves can become important \cite{Crawford}, although this is outside the scope of this paper. In Section \ref{sec3} another competing damping mechanism is presented.\\

\subsection{Coulomb logarithm}\label{sec1.2}
A second important consequence of the RF field is that the Coulomb logarithm $\ln\Lambda$ in Eq. (\ref{1.2}) is modified. This is particularly relevant for UCPs because the traditional expression $\ln\Lambda=\ln\left(\Gamma^{-2/3}\right)$ looses its validity in case of strong coupling $\Gamma\gtrsim 1$. The Coulomb logarithm arises from cutting off the Coulomb collision integral at both large and small impact parameters in elementary calculations of the scattering cross section of an electron by an ion \cite{Mitchner}. However, the physical arguments used to choose these cut-offs are traditionally based on thermal electron velocities only, and the cut-offs will change when in addition the quiver velocity is taken into account. This can be confirmed by explicit calculation \cite{MulserBall}, yielding $\ln\Lambda\approx\ln\left(b_\text{max}/b_\text{min}\right)$, with
\begin{align}
b_\text{max}&=\frac{v_\text{eff}}{\max\left(\omega,\omega_p\right)};\label{1.5a}\\
b_\text{min}&=\frac{e^2}{4\pi\epsilon_0mv_\text{eff}^2};\label{1.5b}\\
v_\text{eff}&\equiv\sqrt{v_\text{th}^2+v_\text{osc}^2}.\label{1.5c}
\end{align}
Here the classical limit $v_\text{eff}<e^2/(2\epsilon_0\hbar)$ has been assumed, where $2\pi\hbar$ is Planck's constant. Eqs. (\ref{1.5a}-\ref{1.5c}) show that also in the Coulomb logarithm, as before, the quiver velocity effectively takes over the role of the thermal velocity in the limit $v_\text{osc}\gg v_\text{th}$. This suggests more generally that kinetic processes in UCPs that depend on the electron temperature may be strongly modified by the presence of an RF field. In the next section, we further validate this notion by showing that the three body recombination rate in an UCP can be strongly suppressed by application of an RF field.

\section{Three-body recombination}\label{sec2}
In the process of three body recombination (TBR), an electron recombines with an ion, while the excess potential energy is carried away by a second electron. In UCPs, TBR is the dominant recombination channel \cite{Killian2} due to the strong scaling of the TBR rate $R$ with temperature, which is $R\propto T_e^{-9/2}$ according to conventional theory \cite{Hinnov,Mansbach}. However, the unphysical divergent behavior of the rate as $T_e\rightarrow0$ indicates that this scaling must break down at sufficiently low temperatures. Modifications of the rate associated with the nonideality of strongly coupled plasmas have been demonstrated analytically \cite{Schlanges,Bornath3,Hahn1,Hahn2} and with molecular dynamics simulations \cite{Bannasch,Pohl}. Also quantum effects associated with the wave character of the electrons can play a role at sufficiently low temperatures, if the electronic De Broglie wavelength becomes noticeable on the spatial scale of the TBR process \cite{Hu2}. On the other hand, in current experiments UCPs remain mainly outside the strongly coupled regime \cite{Robicheaux}, so that numerical models of the expansion dynamics of UCPs that are based on the conventional TBR rate are able to accurately describe experimental results \cite{Gupta}. We will show now that, in addition to any possible strong coupling effects, the presence of an RF field suppresses the TBR rate to a temperature scaling of $R\propto T_e^{-1}$, which is much milder than the conventional $R\propto T_e^{-9/2}$ dependency. We do not consider the mentioned quantum effects, which are presumably small since the quiver motion of the electrons ensures a small De Broglie wavelength.\\

\begin{figure}
\includegraphics[width=\columnwidth]{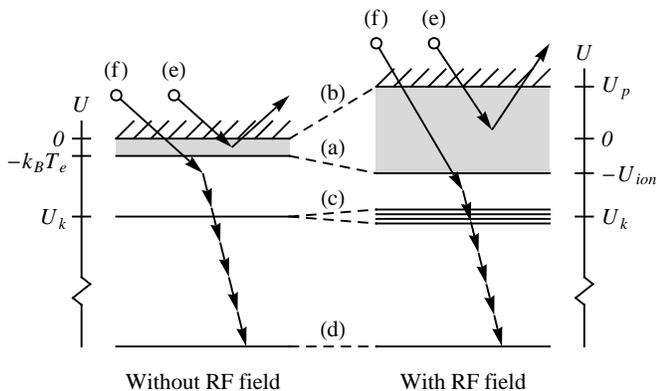}
\caption{\label{fig-1} Energy diagram of three body recombination with and without RF field. The gray bands show energies from which re-ionization is likely. An arbitrary high energy level $U_k<0$ has been drawn; on the sides the energy scale has been indicated. The bottleneck level is moved by the RF field (a). The RF field induces a Stark shift of the continuum threshold (b), Stark splitting of highly excited levels (c); and much smaller shifts of deeply bound states (d). (e): electron that re-ionizes after a collisional transition from the continuum to above the bottleneck level; (f): electron that recombines by making a cascade to deeply bound states after a collisional transition to below the bottleneck level.}
\end{figure}

We determine the TBR rate along the lines of an elementary analytical derivation by Hinnov and Hirschberg \cite{Hinnov}, adapted to the situation in which $v_\text{osc}\gg v_\text{th}$. The TBR rate found by Hinnov and Hirschberg has been confirmed by extensive Monte Carlo simulations \cite{Mansbach} to within a factor of order unity, showing that their model captures the essential physics despite its simplicity. In order to exhibit the RF field effects clearly, we therefore choose to use this simple analytical model rather than performing a detailed numerical study, although the latter will be important to test the results derived here. Let us first briefly review the conventional case where the RF field is absent. Quantum mechanically, a TBR event may be described as an electron making a cascade of transitions between adjacent energy levels of an atom until it reaches the deeply bound states. Under conditions applicable to UCPs, these transitions are mainly effected by collisions with other, free electrons. The process is illustrated in the left panel of Fig. \ref{fig-1}; Considering an electron at any particular energy level $U_i<0$, there is both a finite probability that the next collision will result in an upward transition, and a finite probability that a downward transition results. It can be shown \cite{Hinnov} that the upward transition probability increases with respect to the downward transition probability as $U_i$ grows closer to the continuum, and that upward transitions dominate for levels less than an energy $\sim k_BT_e$ below the continuum. Any electron ending up in the energy band $-k_BT_e<U_i<0$, shown in gray in Fig. \ref{fig-1}, is therefore likely to re-ionize, while electrons below this band are likely to fully recombine. Hence, as far as TBR is concerned, one may qualify the levels $-k_BT_e<U_i<0$ as effectively unbound, and approximate the amount of eventually recombining electrons with those electrons that skip this band altogether by making a direct collisional transition from the continuum to anywhere below the bottleneck level $-k_BT_e$. The validity of this approximation has been confirmed by simulations \cite{Mansbach}. Summing the probabilities of such transitions over all possible initial and final energies of the recombining electron and over all possible energies of the free electron, one finds indeed the usual TBR rate proportional $T_e^{-9/2}$ \cite{Hinnov}.\\

When an RF field is present, two essential modifications must be made to this picture, as illustrated by the right panel of Fig. \ref{fig-1}. First, the RF field interferes with the collisional cascade towards deeply bound levels, because it can ionize electrons from highly excited levels. It is well-known that the character of a field ionization process depends upon the applied field strength and frequency in relation to the binding energy of the electron; accordingly different regimes such as multiphoton- and tunneling ionization may be identified. We consider microwave or lower frequencies and kV/m field strengths, in which case field ionization from highly excited levels is well-described by classical over-the-barrier ionization in a quasistatic electric field \cite{Lankhuijzen}. This has also been verified experimentally \cite{Gallagher2,Gallagher3,Tate}. Accordingly, the combined potential $U=-e/(4\pi\epsilon_0r)-E_0z$ of the ion and the external field has a saddle point along the $z$-axis of height $\sqrt{e^3E_0/(\pi\epsilon_0)}\equiv-U_{ion}$, and any electrons with energies $U_i>-U_{ion}$ will rapidly escape from the ion by going over this saddle point. Such a static description is valid because, in the case at hand, the applied frequency $\omega$ is much smaller than the classical Kepler frequency $\omega_i$ of the energy levels $U_i$ close to $-U_{ion}$. Also the inverse process, in which free electrons enter the vicinity of the ion in the presence of a low-frequency field and which is the low-frequency equivalent of stimulated radiative recombination, has been observed \cite{Shuman,Overstreet}. The lowering of the Coulomb barrier to $-U_{ion}$ due to the external field thus defines a range of energies $U>-U_{ion}$ that are effectively unbound. Regarding the three body recombination process, any electron ending up in this energy range is more likely to ionize than to proceed with a downward collisional cascade. Thus, analogous to the field-free case, only free electrons that make a direct collisional transition to states below the bottleneck level $-U_{ion}$ will contribute to the TBR rate, but now the bottleneck level is set by the field and no longer by the plasma property $-k_BT_e$.\\

A second influence of the RF field is the fact that the energy of both free and bound electrons will change due to the field. For free electrons, the energy increment is just the quiver energy $U_p=mv_\text{osc}^2/4$. As a result the continuum threshold shifts up by $U_p$ as well (see Fig. \ref{fig-1}), which is a well-known effect in multiphoton ionization experiments \cite{Bucksbaum}. This upshift is important for the TBR process since free electrons will now have to loose an additional energy $U_p$ in order to recombine with an ion. Combined with the adapted bottleneck, the minimum energy loss to effect a TBR event has thus increased from $k_BT_e$ in the field-free case to $U_p+U_{ion}$ in the case with field, as is illustrated in Fig. \ref{fig-1} by the gray bands. This suppresses the TBR rate significantly. Finally, the energy change of the bound levels due to the RF field is the AC Stark shift. However, the energy levels that are available for TBR are the levels below $-U_{ion}$, for which the shift is approximately equal to the DC Stark shift because $\omega_i\gg\omega$. For states just below $-U_{ion}$, the electric field exceeds the Inglis-Teller limit, which means that the Stark splitting of the manifolds with principal quantum number $k$ is large enough to fill the energy space with states more or less homogeneously \cite{Lankhuijzen}. An additional observed effect due to this strong Stark mixing in an AC field is that electrons may ionize from below $-U_{ion}$ via subsequent upward Landau-Zener transitions \cite{Lankhuijzen}. We neglect this effect because it is a much slower process than direct over-the-barrier-ionization \cite{Watkins}. Resonant atomic transitions that might be induced by the RF field are not included either, although they may have an effect on the collisional cascade.\\

We now recalculate the TBR rate in the presence of an RF field, taking account of the field modifications described above. By the method of detailed balance, under the hypothetical condition of thermal equilibrium the rate of collisional transitions from the continuum $U>U_p$ to the bound energy level $U_i<U_{ion}$ is equal to the rate of the inverse process, which are ionizing transitions from the bound level to the continuum caused by electron impact. From the well-known \cite{Hinnov} cross section $S_i(U)$ for a collisional energy transfer of at least $\left|U_i\right|+U_p$ from a moving electron with energy $U$ to a stationary electron, the rate of collisional ionization from level $U_i$ per unit plasma volume is
\begin{align}
R_i=\int_{\left|U_i\right|+U_p}^\infty n_in_evS_i(U)f(U)dU.\label{1.6}
\end{align}
Here, $f(U)$ is the energy distribution function of free electrons, $v$ is the electron velocity corresponding to energy $U$, and $n_i$ is the density of electrons in level $U_i$. The rate of TBR via level $i$, which is the inverse process, is obtained by substituting for $n_i$ the equilibrium value from the Saha equation \cite{Mitchner}, because the two rates must be equal at equilibrium. Let us first consider the case where $U_p\ll U_{ion}$, that is, for relatively high frequencies or low fields, and denote the corresponding TBR rate by $R_0$. In this case $U\gg U_p$ in the whole integration domain of Eq. (\ref{1.6}), so that $f(U)$ may be approximated by an ordinary Boltzmann distribution without the need to correct it for the quiver motion. Evaluating the integral in Eq. (\ref{1.6}), substituting the Saha value for $n_i$, and summing over all energy levels below $-U_{ion}$, gives the total TBR rate
\begin{align}
R_0=\!\sum_{i}R_i\approx\!\frac{e^4\hbar^3n_e^3}{4g\epsilon_0^2m^2\left(k_BT_e\right)^3}\!\int_{-\infty}^{-U_{ion}}\!\!\!\!\!\!F\!\left(\frac{U}{k_BT_e}\right)\!D(U)dU,\label{1.7}
\end{align}
where $F(x)\equiv\exp(-x)\Ei(x)-1/x$ with $\Ei$ the exponential integral \cite{Abramowitz} and $g$ is the degeneracy of the ionic ground state. The sum over states has been approximated by an integral over the bound energy, resulting in the density of states $D(U)$ as a factor. Approximating the atomic potential with that of hydrogen, $D(U)\approx\nicefrac{1}{2}\Ry^{3/2}\left|U\right|^{-5/2}$, where $\Ry=13.6$ eV is the Rydberg energy. For kV/m field strengths, $\left|U\right|/(k_BT_e)\gg1$ over the whole integration domain of Eq. (\ref{1.7}), so that the function $F$ can be approximated by its asymptotic value $F\approx\left(k_BT_e/U\right)^2$. Then the remaining integral contains the field effects, but is independent of the temperature. This means that the temperature scaling of the TBR rate that is derived here is insensitive to errors due to our approximate description of the energy $U_{ion}$ and the Stark shift structure, although the prefactor may change somewhat in a more detailed calculation. Integration of Eq. (\ref{1.7}) gives
\begin{align}
R_0&\approx\frac{\pi^2}{7g}\sqrt{\frac{2}{m}}\left(\frac{e^2}{4\pi\epsilon_0}\right)^5\frac{n_e^3}{U_{ion}^{7/2}k_BT_e}\approx\nonumber\\
&\approx2.6\cdot10^{-27}\frac{n_e^3[\text{cm}^{-9}]}{U_{ion}^{7/2}k_BT_e[\text{eV}^{9/2}]}\text{cm}^{-3}\text{s}^{-1},\label{1.8}
\end{align}
assuming $g=2$. Within a factor of order unity, this three body recombination rate is equal to the accepted result for the case without RF field \cite{Mansbach}, except that $7/2$ powers of $k_BT_e$ have been replaced an energy $U_{ion}$ characterizing the applied field. This reduces the strongly divergent behavior $R\propto T_e^{-9/2}$ to the much milder dependency $R\propto T_e^{-1}$. Thus three body recombination may be significantly suppressed by the application of an RF field. A similar electric-field induced suppression of the TBR rate has been considered before \cite{Hahn1}, although in that work the plasma microfield or Holtsmark field was taken into account rather than an externally applied field. The calculated TBR rate for singly charged ions was $1.4\cdot10^{-31}\Gamma_Z n_p^7n_e^3/(k_BT_e)$ in the units of Eq. (\ref{1.8}), with $\Gamma_Z\approx2$ and $n_p$ the principal quantum number at their bottleneck level defined in the paper. Using our bottleneck level instead by substituting $n_p=\sqrt{\Ry/U_{ion}}$ precisely gives Eq. ({\ref{1.8}), including the correct numerical factor, showing that both results are in agreement.\\

Eq. (\ref{1.8}) is valid for $U_p\ll U_{ion}$ only. However, the calculation is easily generalized to arbitrary $U_p$, the only added complication being the need to include the quiver motion of the free electrons. The details are given in Appendix \ref{secC}; the result is
\begin{align}
R=R_0G\left(\frac{U_p}{U_{ion}}\right),\label{1.9}
\end{align}
where $R_0$ is the rate given by Eq. (\ref{1.8}) and $G$ is a correction factor. The latter is given by Eq. (\ref{C4}) and is approximately equal to
\begin{align}
G(x)\approx\left[1+\left(\beta x\right)^{1/\alpha}\right]^{-5\alpha/2},\label{1.10}
\end{align}
with $\alpha=1.137$ and $\beta=(2/7)^{2/5}$.

\section{Collisionless absorption}\label{sec3}
\subsection{Absorption models}\label{sec3.0}
Even without the presence of electron-ion collisions, individual electrons in a plasma can absorb energy from an applied electric field. For bulk plasmas, this collisionless absorption effect is the well-known Landau damping \cite{Spitzer}, in which electrons can gain net energy from a high-frequency propagating electric wave, despite the fact that the high-frequency electric force tends to cancel out on the average. This is possible when the thermal velocity of the electron is close to the velocity of the wave, so that the electric field is approximately static in the electron frame of reference. Essential for this mechanism is a resonance between thermal motion and applied field. In plasmas of finite size, such as an UCP, the thermal motion of electrons is necessarily confined by the plasma boundaries, so the assumption of rectilinear motion implicit in the Landau damping mechanism of bulk plasmas is no longer appropriate. Rather, the electrons perform quasi-periodic motion in the electrostatic potential of the plasma, as is detailed below. Furthermore, the electric field in the plasma is homogeneous rather than a propagating wave when the applied wavelength is much larger than the plasma size, such as in the case of an RF field applied to an UCP. Nevertheless, electrons may on the average gain energy, and this is again due to a resonance between the thermal motion and the applied field. This is why the collisionless absorption of finite plasmas has been called Landau damping as well \cite{Yannouleas,Zaretsky2}, although the character of the correlation is quite different. In this section, we calculate the RF energy absorption of an UCP by this mechanism. To avoid confusion, it should be noted that the resonance between thermal motion and RF field that is meant here has nothing to do with the more familiar plasma resonance. The electrons in the plasma have an individual thermal motion superposed on a collective quiver motion; the resonance meant here concerns the first of these, while plasma resonance relates to the latter.\\

First, we mention a number of other approaches to collisionless absorption and argue why these are less appropriate for UCPs in RF fields. In the above description of collisionless absorption, the applied field plays the role of a perturbation on the thermal motion of the electrons. One may change perspective and look at the quiver motion of the electrons as being the primary motion, perturbed by a thermal one. Because the details of the thermal motion are determined by the details of the plasma potential, this can be interpreted as an oscillating electron having interaction with the plasma potential itself. This view is particularly appropriate when the potential can be approximated by an infinitely deep well, so that the 'interaction with the potential' simply becomes 'collisions with the plasma boundary'. Then the collision frequency of electrons with the plasma boundary is on average
\begin{align}
\nu_\text{p}\sim\frac{v}{\sigma}\hspace{1cm}\text{(hard wall model)},\label{3.0}
\end{align}
where $\sigma$ is the plasma size, and $v$ is the characteristic velocity of the electrons that is taken to be the thermal velocity \cite{Zaretsky}, a combination of thermal and quiver velocity \cite{Megi} or Fermi velocity \cite{Yannouleas} depending on the model used. On average the electrons gain an energy $2U_p$ per hard wall collision, in analogy with Eq. (\ref{1.1}). The result (\ref{3.0}) also follows as a special case from the more general Landau damping approach when specialized to a hard wall potential \cite{Zaretsky}. While a flat potential with hard walls, and hence the resulting absorption rate $2vU_p/\sigma$, may be a good approximation for large metallic clusters \cite{Yannouleas, Megi}, it is not for UCPs. In the process of creation of an UCP from an atomic cloud, part of the electrons escape from the plasma immediately after photoionization of the cloud. This continues until the accumulated charge imbalance self-limits further loss of electrons. The resulting spherically symmetric Coulomb potential of the UCP with a typical Gaussian density distribution is \cite{Killian2}
\begin{align}
U(r)=U_0\left[1-\frac{\sqrt{\pi}\sigma}{2r}\erf\left(\frac{r}{\sigma}\right)\right],\label{3.1}
\end{align}
where $\erf(r/\sigma)$ denotes the error function \cite{Abramowitz}, and $r$ is the distance to the cloud center. The depth of the potential saturates to $U_0\sim k_BT_e$ by nature of the charging process. Clearly, the hard wall potential is not a very good approximation in this case and a more detailed calculation of the energy absorption is necessary to account for the smoothness of the potential.\\

Another absorption mechanism that is considered important for large metal clusters is the Brunel effect \cite{Brunel}, in which electrons at the plasma boundary are pulled out of the plasma by the applied electric field and then driven back into the plasma as the field reverses direction. When the plasma is sufficiently overdense, the interaction effectively stops once the electron has moved deeper into the plasma than the skin depth, resulting in net energy gain because the electron cannot be brought back to rest by the evanescent field. The resulting absorption rate, divided by $2U_p$ for comparison, gives again the hard wall collision frequency Eq. (\ref{3.0}), with $v$ the high-frequency velocity. In an UCP, however, the Brunel mechanism is not in effect either, since typically the skin depth, which is comparable to $c/\omega_p$ with $\omega_p$ the plasma frequency, is much larger than the plasma size.\\

Finally, when the applied field is so strong that the oscillation amplitude of individual electrons is comparable to or larger than the plasma size, one can hardly speak of the applied field as a perturbation, and other descriptions of the electron motion such as nonlinear oscillators \cite{MulserOsc,Kundu,Kostyukov} or scattering off the plasma potential \cite{Kostyukov2} are more appropriate. Here we do not consider such strong field effects.\\

\subsection{RF absorption by electrons in a general potential}\label{sec3.2}
We now proceed to calculate the collisionless RF energy absorption by an UCP, taking account of the smooth plasma potential shape shown in Eq. (\ref{3.1}) rather than resorting to a hard wall approximation. We make use of the calculational method developed by Zaretsky et al. \cite{Zaretsky}. When forcing an UCP with an RF signal, the electric field in the plasma consists of the external RF field, the polarization field caused by any excited plasma modes, and the field corresponding to the plasma potential Eq. (\ref{3.1}). The combination of the first two fields may be considered a fast harmonic perturbation on the latter field. Although UCPs behave entirely classically \cite{Killian2}, a quantum mechanical description of this situation proves best suited to calculate the RF energy absorption. Accordingly, the electrons occupy bound states in the plasma potential, and can change states by absorption or emission of an RF photon. The quantum mechanical calculation of the absorption is given in detail in Ref. \cite{Zaretsky}. A spatially homogeneous RF field is assumed, which rules out strong local field enhancements such as those generated by plasma resonances. Therefore the following calculation is restricted to underdense plasmas. In summary, perturbation theory is applied, in which the transition probability of electrons between any pair of states is given by Fermi's Golden rule \cite{Landau}. The number of RF photons absorbed by the plasma equals the difference between the number of electron transitions to a higher state and those to a lower state, and the absorbed RF energy is this amount multiplied by the photon energy. Exploiting, in addition, the fact that the system dimension $\sigma$ is much larger than the typical De Broglie wavelength of the electrons, one can adopt the quasi-classical or Bohr-Sommerfeld theory to approximate quantum mechanical quantities by their classical analogues \cite{Landau}. Although results for a general three-dimensional potential are available \cite{Zaretsky}, we will use the one-dimensional analogs because then the mathematics is much more transparent. This does not represent a major error since the energy transfer from the RF field to the plasma proceeds via electrons that move partially resonant with the applied field. This means that only one component of the electron trajectories, namely the one that is parallel to the applied field, contributes to the RF absorption, so that the problem is essentially one-dimensional. Explicit calculation of the RF absorption in both the full three-dimensional and corresponding one-dimensional hard wall potential \cite{Zaretsky} confirms that the latter captures the general behavior.\\

Expressing as before the absorbed RF power $P_p$ due to collisionless absorption in terms of an effective frequency $\nu_\text{p}$, it is found that \cite{Zaretsky}
\begin{align}
P_p&=2\nu_pU_p;\label{3.6a}\\
\nu_\text{p}&=\frac{\pi m\omega^3}{Zk_BT_e}\sum_{s=0}^\infty \left[\frac{\left|X(\epsilon)\right|^2}{\left|d\Omega/d\epsilon\right|}\exp\left(-\frac{\epsilon}{k_BT_e}\right)\right]_{\epsilon=\epsilon_s}.\label{3.6}
\end{align}
Here, $\Omega(\epsilon)$ is the oscillation frequency of the classical trajectory $x(\epsilon,t)$ of a particle with energy $\epsilon$ in the unperturbed potential,
\begin{align}
X(\epsilon)=\frac{\Omega(\epsilon)}{2\pi}\int_0^{2\pi/\Omega(\epsilon)}x(\epsilon,t)\exp\left(i\omega t\right) dt\label{3.7}
\end{align}
is the Fourier component of the classical trajectory at the frequency of the perturbation,
\begin{align}
Z=\int\exp\left(-\frac{\epsilon}{k_BT_e}\right)\frac{d\epsilon}{\Omega(\epsilon)}\label{3.8}
\end{align}
is the partition function of the electron distribution over the energy states, which is assumed a Boltzmann distribution here, and the sum in Eq. (\ref{3.6}) is over energies that are roots of the equation
\begin{align}
\left(2s+1\right)\Omega(\epsilon_s)=\omega.\label{3.9}
\end{align}
Without attempting to explain all details underlying Eqs. (\ref{3.6}-\ref{3.9}) here, it is noted \cite{Zaretsky} that the only contributions to the absorbed energy come from those electrons whose trajectory is in resonance with the applied field according to Eq. (\ref{3.9}). This is the correlation between thermal motion and applied field also characteristic for bulk Landau damping. Furthermore, the contributions in Eq. (\ref{3.6}) are proportional to $\left|X\right|^2$, the spectral content of the trajectory at the applied frequency. However, the dominant frequencies in the spectrum of the trajectory will be on the order of the oscillation frequency $\Omega(\epsilon)$. In a potential such as Eq. (\ref{3.1}) with $r$ replaced by $x$, this frequency will be comparable to that of a harmonic oscillator potential with the same curvature at $x=0$, that is, to $\Omega\sim\sqrt{2U_0/(3m\sigma^2)}\equiv\omega_0$. Therefore, it is expected that the RF energy absorption strongly depends on the ratio $\omega/\omega_0$. In addition, the ratio of particle energy $\epsilon$ and thermal energy $k_BT_e$ appears in Eq. (\ref{3.6}), the former being limited to values smaller than the potential depth $U_0$, so there will be some weak secondary dependency on the ratio $U_0/(k_BT_e)$ as well. These properties are indeed found below.\\

In the classical UCP system the spacing between energy levels is much smaller than the thermal energy, therefore the sum in Eq. (\ref{3.6}) may be approximated by integration over $s$. A subsequent change of integration variable from $s$ to the energy $\epsilon_s$ introduces an extra factor $(d\epsilon_s/ds)^{-1}$, which is the density of resonant states. This factor is obtained by differentiating Eq. (\ref{3.9}) with respect to $s$, yielding $\left|d\Omega/d\epsilon\right|_{\epsilon=\epsilon_s}\cdot d\epsilon_s/ds=2\Omega^2/\omega$. Accordingly, Eq. (\ref{3.6}) becomes
\begin{align}
\nu_{p}\approx\frac{\pi m\omega^4}{2Zk_BT_e}\int\left|\frac{X(\epsilon)}{\Omega(\epsilon)}\right|^2\exp\left(-\frac{\epsilon}{k_BT_e}\right)d\epsilon,\label{3.10}
\end{align}
where the subscript $s$ has been dropped.\\

\begin{figure}
\includegraphics[width=\columnwidth]{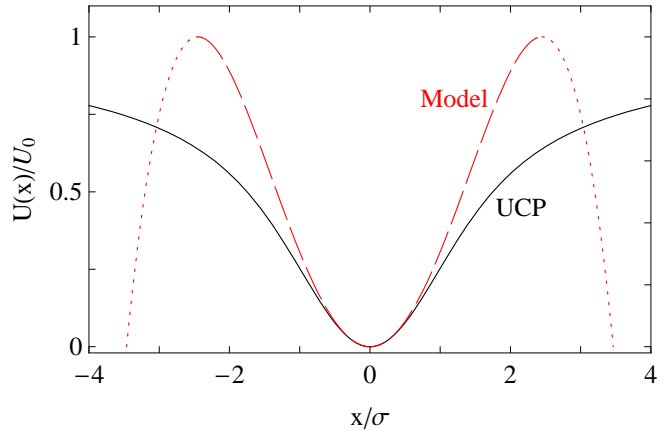}
\caption{\label{fig0} (Color online) Model potential (red dashed line, Eq. (\ref{3.2})) compared to the actual UCP potential (black solid line, Eq. (\ref{3.1})). The parameters have been set to $\omega_1=\omega_0$ and $a$ such that $U_1=U_0$. The dotted parts of the model potential are not used.}
\end{figure}

\subsection{RF absorption in a model plasma potential}\label{sec3.1}
Eq. (\ref{3.10}) allows explicit calculation of the absorbed RF power, if the classical trajectories in the potential are known analytically. However, for the particular potential Eq. (\ref{3.1}), closed expressions for the trajectories are not available. In order to still make quantitative estimates for the energy absorption, instead of Eq. (\ref{3.1}) we use a model potential with the same general shape for which the trajectories are known analytically:
\begin{align}
U(x)=\frac{m\omega_1^2x^2}{2}\left(1-\frac{x^2}{a^2}\right),\label{3.2}
\end{align}
where $a$ is a positive constant with units of length. Eq. (\ref{3.2}) is the potential of a Duffing oscillator commonly used to describe the motion of a mass on a cubic softening spring. Although this potential differs from the actual UCP potential Eq. (\ref{3.1}), we note that from a physical point of view the most important characteristics of the UCP potential are the temperature, which sets the potential depth $U_0$, and the charge density, which sets the curvature $m\omega_0^2$ at the bottom of the potential. Therefore we should obtain a reasonable estimate for the energy absorption by choosing the model potential accordingly, setting the curvature $m\omega_1^2$ equal to $m\omega_0^2$ and the potential depth $m\omega_1^2a^2/8\equiv U_1$ equal to $U_0$. Important as well is that the infinitely differentiable UCP potential is modeled by an equally smooth one, and that both potentials approach their edge with vanishing slope. In Fig. \ref{fig0} the two potentials are compared. \\

A particle is bound by the potential Eq. (\ref{3.2}) only if its energy $\epsilon$ is less than $U_1$. For such a bound particle the classical trajectory, starting at time $t=0$ at position $x=0$, can be shown to be given by the periodic function \cite{Lawden}
\begin{align}
x(\epsilon,t)=a\sqrt{\frac{u}{2v}}\sn\left(\sqrt{\frac{v}{2}}\omega_1 t,\frac{u}{v^2}\right),\label{3.3}
\end{align}
where $\sn(y,m^2)$ is the Jacobi elliptic function with argument $y$ and modulus $m$, and $u=\epsilon/U_1$ is the particle energy in units of the potential depth, and $v=1+\sqrt{1-u}$. The frequency $\Omega$ with which the particle oscillates back and forth in the potential is given by \cite{Lawden}
\begin{align}
\Omega(\epsilon)=\frac{\pi\sqrt{v}}{2\sqrt{2}\K\left(u/v^2\right)}\omega_1,\label{3.4}
\end{align}
where $\K(m^2)$ is the complete elliptic integral of the first kind with modulus $m$. In the limit of vanishing particle energy $\epsilon\rightarrow0$, the trajectory (\ref{3.3}) approaches harmonic motion with frequency $\omega_1$, while the motion becomes anharmonic with the frequency monotonically decreasing to zero as the energy grows to $U_1$.\\

In Appendix \ref{secA} the absorbed power is calculated by using Eqs. (\ref{3.3}) and (\ref{3.4}) in Eq. (\ref{3.10}). The exact result Eq. (\ref{A2}) for the effective collision frequency is plotted in Fig. \ref{fig1} as a function of $\omega_1/\omega$, assuming a potential depth equal to $k_BT_e$. Also plotted is the asymptotic approximation, valid for $\omega_1/\omega\ll1$,
\begin{align}
\left(\frac{\nu_\text{p}}{\omega}\right)_\text{Model}=C\left(Y\right)\left(\frac{\omega}{\omega_1}\right)^2\exp\left(-\sqrt{2}\pi\frac{\omega}{\omega_1}\right),\label{3.11}
\end{align}
which fits the exact result very well. In a typical UCP, $\sigma\sim1$ mm and $T_e\sim1$ K \cite{Killian2}, while in a typical RF experiment $\omega/(2\pi)>1$ MHz \cite{Kulin}, so that usually the asymptotic regime of Eq. (\ref{3.11}) is in effect. The prefactor $C\left(Y\right)$ is given by Eq. (\ref{A3}) and depends on the ratio $Y=U_1/k_BT_e$. As argued previously, the choice of model potential parameters that best represents the actual UCP potential is $\omega_1=\omega_0$ and $U_1=U_0\sim k_BT_e$~, giving $Y\sim1$. The corresponding prefactor in Eq. (\ref{3.11}) lies in the range $C=20-35$ for $Y=0.5-2.0$.\\

\begin{figure}
\includegraphics[width=\columnwidth]{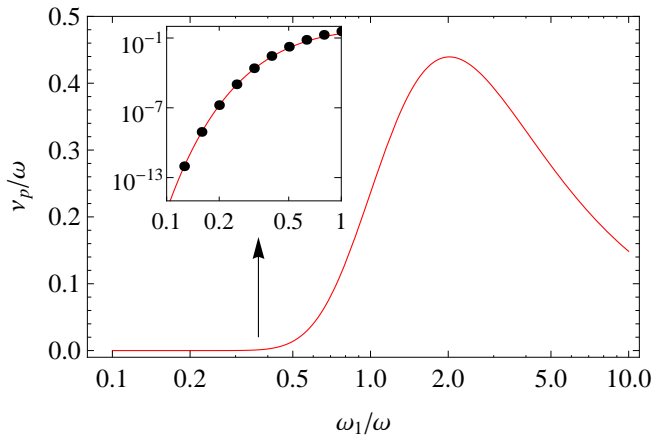}
\caption{\label{fig1} (Color online) Effective collision frequency $\nu_\text{p}$ for RF absorption due to the finite plasma size, as a function of the ratio of the frequency characterizing the potential $\omega_1$ to the RF frequency $\omega$. A potential depth equal to $k_BT_e$ has been assumed. The inset shows the behavior for $\omega_1/\omega<1$ on a logarithmic scale, comparing the exact result Eq. (\ref{A2}) (solid curve) to the approximate result Eq. (\ref{3.11}) (dots).}
\end{figure}

From Fig. \ref{fig1} and Eq. (\ref{3.11}), it is clear that the collisionless RF absorption by an UCP strongly depends on $\omega_1/\omega$, that is, on the ratio of the frequency at which the thermal motion of the UCP electrons takes place to the RF frequency. This strong dependency was anticipated above from the fact that the collision frequency Eq. (\ref{3.6}) is proportional to the spectral content of the trajectory at the RF frequency: when $\omega_1$ and $\omega$ do not differ too much, the RF forcing and the electron motion take place on more or less the same time scale, so that the electron motion contains an appreciable Fourier component at the RF frequency, resulting in resonant and efficient energy transfer. Since all oscillation frequencies given by Eq. (\ref{3.4}) are in fact less than $\omega_1$, the average oscillation frequency will be less than $\omega_1$ as well, so that in Fig. \ref{fig1} the peak in the energy transfer occurs at a somewhat higher value than $\omega_1/\omega=1$, corresponding to a somewhat slower forcing.\\

An important feature of Fig. \ref{fig1} and Eq. (\ref{3.11}) is the threshold-like behavior of $\nu_p$: for $\omega_1\gtrsim\omega$ the absorption is significant, while for $\omega_1/\omega\rightarrow0$ it decreases exponentially. The inset shows that this decrease is very rapid, so that collisionless absorption is completely negligible if $\omega_1\ll\omega$. This condition can be written as $1\gg\omega_1/\omega\sim\omega_0/\omega\equiv\sqrt{2U_0/(3m\sigma^2\omega^2)}\sim v_\text{th}/(\sigma\omega)$. Physically, this corresponds to the situation in which a low temperature yields by assumption a shallow potential with slow electrons, so that almost no electrons traverse the plasma within one RF oscillation. Combined with the lack of steep features in the smooth potential, this means that there is almost no electron motion available at the RF frequency that is susceptible to resonant absorption. One may thus define a critical temperature
\begin{align}
k_BT_p=m\omega^2\sigma^2\label{3.11a}
\end{align}
that separates a temperature regime $T_e\gtrsim T_p$ in which collisionless absorption is significant and a regime $T_e\ll T_p$ where it is negligibly small. Note that this behavior is not at all described by the hard wall approximation Eq. (\ref{3.0}). The reason for this is that an electron bouncing between hard plasma boundaries abruptly changes its velocity at every wall collision, giving rise to high-frequency components essentially regardless of the velocity. Therefore Eq. (\ref{3.0}) predicts significant collisionless absorption at any temperature, but is valid only for steep plasma potentials.

\subsection{Validity for the actual UCP potential}\label{sec3.3}
As we just described, the collisionless absorption rate in the model plasma potential exponentially decreases with the ratio $v_\text{th}/(\sigma\omega)$. Since the physical arguments leading to Eq. (\ref{3.11a}) are valid for any general smooth plasma potential, also in actual UCPs the collisionless absorption rate will quickly decrease once the electron temperature is below the critical temperature $T_p$. However, one may still ask whether the decay constant of this decrease (i.e. the factor $\sqrt{2}\pi$ in Eq. (\ref{3.11})) is also representative for actual UCPs, or depends on the potential shape. Lacking analytical expressions for the trajectories $x(t)$ in the UCP potential, this cannot be verified by explicit calculation. Nevertheless, the decay constant can be calculated by quantifying the asymptotic behavior of the Fourier coefficients of the trajectories, using the so-called Darboux's Principle \cite{Boyd}. This however requires considering the analytical continuation of $x(t)$ to the complex $t$-plane. The details are rather technical and are relegated to Appendix \ref{secB}. The main result is that the quantity $\left|X(\epsilon)\right|^2$ in Eq. (\ref{3.10}) for the UCP potential contains an extra factor of approximately $\exp\left(2\omega/\omega_0\right)$ as compared to the case of the model potential, independent of the particle energy $\epsilon$ and for sufficiently large $\omega/\omega_0$. Including this extra factor in the result Eq. (\ref{3.11}), the asymptotic rate of decrease of the collision frequency is approximately equal to
\begin{align}
\left(\frac{\nu_\text{p}}{\omega}\right)_\text{UCP}\propto\exp\left[-\left(\sqrt{2}\pi-2\right)\frac{\omega}{\omega_0}\right].\label{3.12}
\end{align}
Although the decay constant $\sqrt{2}\pi-2$ is smaller than that of Eq. (\ref{3.11}) and Fig. \ref{fig1}, it is still of the same order of magnitude. Also in the UCP case, therefore, the collisionless absorption is negligible for $\omega_0\ll\omega$, or equivalently for temperatures below $T_p$ given by Eq. (\ref{3.11a}).

\section{Discussion and conclusions}\label{sec4}
In this paper, we considered three mechanisms by which an RF field influences the temperature of an UCP. First, RF energy is absorbed through the well-known process of collisional absorption, in which electrons gain energy during Coulomb collisions with ions. Second, the RF field modifies the TBR rate by ionizing electrons from intermediate high-lying Rydberg states. Third, resonance between the motion of electrons in the plasma potential and the RF field may give rise to collisionless energy absorption. For all of these processes, na\"ive extrapolations from well-known formulas are inadequate for UCPs or strong RF fields. For example, the electron-ion collision frequency Eq. (\ref{1.3}) is much smaller than the Spitzer frequency for strong RF fields, suppressing the collisional absorption rate. As we indicated, this is because the quiver velocity effectively takes over the role of the thermal velocity, or equivalently, because the temperature is replaced by the ponderomotive potential in the collision frequency. Likewise, the TBR rate in strong RF fields is much smaller than expected from the commonly used $T_e^{-9/2}$-scaling, partly because the conventional TBR bottleneck level characterizing the plasma is replaced by the energy $U_{ion}$ characterizing the RF field. Figure \ref{fig7} schematically shows the various heating regimes in terms of the RF field amplitude and frequency; the strong-field effects apply to the area above the slanted line. As discussed in the previous section, collisionless absorption is only relevant at sufficiently high temperatures or low frequencies, as is represented by the area to the left of the vertical line in Fig. \ref{fig7}.\\

\begin{figure}
\includegraphics[width=0.9\columnwidth]{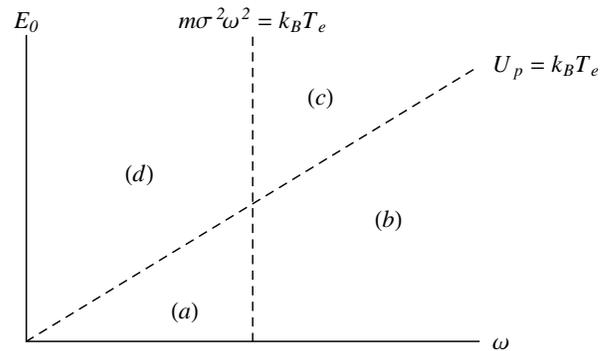}
\caption{\label{fig7} Heating regimes for RF-driven UCPs in terms of the applied frequency $\omega$ and field strength $E_0$. (a,b): Collisional absorption rate according to Spitzer collision frequency and TBR rate according to $T_e^{-9/2}$-scaling; (c,d): Collisional absorption rate according to collision frequency Eq. (\ref{1.3}) and TBR rate according to Eq. (\ref{1.8}). (a,d): Collisionless absorption relevant; (b,c): Collisionless absorption negligible.}
\end{figure}

Let us conclude by giving two numerical examples. The RF experiment of Fletcher et al. \cite{Fletcher} was well in the weak-field regime (\emph{a,b}) of Fig. \ref{fig7} according to the reported experimental values. Using these values in Eqs. (\ref{1.1}-\ref{1.5c}), (\ref{3.6a}) and (\ref{3.11}) gives absorption rates per electron of $P_\text{ei}/k_B=3\text{ K}/\mu\text{s}$ and $P_p/k_B=0.002\text{ K}/\mu\text{s}$ at the highest reported frequency and amplitude. Considering the electron temperature of 100 K and the typical plasma expansion time of microseconds, these low absorption rates will not influence the plasma temperature and expansion much. For somewhat larger RF amplitudes, however, the collisional absorption starts to become significant on the time scale of the plasma expansion, which may be related to the high-field effects observed in the experiment.\\

As an example in the regime (\emph{c}) of Fig. (\ref{fig7}), consider an applied field with an amplitude of 0.1 MV/m at a frequency of 28 GHz, which is currently available \cite{Kariya}. We deliberately choose this relatively high frequency because otherwise the oscillation amplitude of the plasma electrons would exceed the plasma size at such a large field strength, which situation is outside the scope of this paper. Choosing further $\sigma=1$ mm, $T_e=1$ K and $n=10^{8}$ cm$^{-3}$, Eqs. (\ref{1.1}-\ref{1.5c}), (\ref{1.9}),(\ref{3.6a}) and (\ref{3.11}) give $P_\text{ei}/k_B=4\cdot10^2\text{ K}/\mu\text{s}$ and $R/n_e=3\cdot10^{-7}$ $\mu$s$^{-1}$, while the collisionless absorption rate is vanishingly small. Thus collisional absorption is expected to heat the plasma to the 100 K scale during the expansion time of the plasma, while the chance that an individual electron recombines is very small. Now compare these numbers to the corresponding results obtained from standard expressions. Using the Spitzer collision frequency instead of Eq. (\ref{1.3}) would give $P_\text{ei}/k_B=4\cdot10^5\text{ K}/\mu\text{s}$, which would predict immediate heating of the UCP to conventional eV plasma temperatures. According to the usual $T_e^{-9/2}$-scaling (Eq. (\ref{1.8}) with $U_{ion}$ replaced by $k_BT_e$), the TBR rate per electron would be $R/n_e=50$ $\mu$s$^{-1}$. Assuming an energy release of $\sim k_BT_e$ per recombination, this would result in a heating rate per electron on the order of $10^2$ K/$\mu$s due to TBR alone, although of course this rate would be quickly quenched as the electron temperature rises. Based on the hard wall approximation Eq. (\ref{3.0}) with $v=v_\text{th}$, the collisionless absorption rate would be $P_p/k_B=1\cdot10^3$ K/$\mu$s rather than exponentially small. From these numbers it is clear that it is essential to properly take into account strong field effects on the one hand, and the smooth UCP plasma potential on the other hand. For the application of a very strong microwave field to an UCP, it changes the predicted effect from destroying the plasma immediately to only heating it up moderately.\\

In summary, we have analytically studied well-known plasma heating mechanisms and specialized them to the system of an UCP driven by a uniform, and possibly strong, RF field. Benchmarking our results against molecular dynamics simulations will yield valuable additional insights, and will also identify any additional RF effects that are not addressed in this paper. Among these are, for example, plasma cloud deformations expected when the electron oscillation amplitude becomes comparable to the plasma size, relativistic effects, plasma waves and other instabilities. Experiments in which RF fields are used to probe plasma resonances rely on adequate modeling of the UCP expansion dynamics, which will benefit from detailed knowledge of RF heating mechanisms such as those discussed in this paper. Furthermore, in virtue of comparable coupling parameters, RF-driven UCPs may be seen as millimetre-sized scale models of laser-driven solid state density plasmas. Understanding the ways in which ultracold plasmas interact with RF fields is therefore also relevant for such high-density systems.\\

\begin{acknowledgments}
We would like to thank R.M.W. van Bijnen for helpful discussions.
\end{acknowledgments}

\appendix
\section{Effective collision frequency}\label{secA}
The Fourier series of the trajectory (\ref{3.3}) equals \cite{Lawden}
\begin{align}
x(\epsilon,t)=2a\frac{\Omega}{\omega_1}\sum_{n=0}^\infty\frac{\sin\left[\left(2n+1\right)\Omega t\right]}{\sinh\left[\left(2n+1\right)\frac{\pi\K\left(1-u/v^2\right)}{2\K\left(u/v^2\right)}\right]}.\label{A0}
\end{align}
Substituting Eq. (\ref{3.9}) in Eq. (\ref{3.7}), and comparing with (\ref{A0}), it follows that
\begin{align}
\left|\frac{X(\epsilon)}{\Omega(\epsilon)}\right|=\frac{a}{\omega_1}\csch\left[\frac{\omega}{\omega_1}\sqrt{\frac{2}{v}}\K\left(1-u/v^2\right)\right].\label{A1}
\end{align}
Using this quantity in Eq. (\ref{3.10}), and changing the integration variable to $u=\epsilon/U_1$, results in
\begin{align}
\frac{\nu_\text{pot}}{\omega}=&\pi^2\sqrt{2}\left(\frac{\omega}{\omega_1}\right)^3\cdot\label{A2}\\
&\cdot Y\frac{\int_0^1\csch^2\left[\frac{\omega}{\omega_1}\sqrt{\frac{2}{v}}\K\left(1-u/v^2\right)\right]\exp\left(-Yu\right)du}{\int_0^1v^{-1/2}\K\left(u/v^2\right)\exp\left(-Yu\right)du}\nonumber,
\end{align}
where $Y=U_1/k_BT_e$. The integrations are over energies smaller than the potential depth, corresponding to bound electrons, since transitions to the continuum do not give rise to energy increase of the ensemble that is left behind. When $\omega_1/\omega\ll1$, to a good approximation $\csch Z\approx2\exp\left(-Z\right)$ in the numerator of Eq. (\ref{A2}). Furthermore, the argument $Z$ of the $\csch$-function is smallest at $u=1$, so that the region around the upper integration limit will give the dominant contribution to the integral in Eq. (\ref{A2}), and $Z$ may be approximated by its Taylor series around $u=1$. This gives $\csch Z\approx2\exp\left[-(\pi/\sqrt{2})(\omega/\omega_1)\left(1+3\delta/16\right)\right]$, where $\delta=1-u$. Similarly, in the integral in the denominator of Eq. (\ref{A2}), the elliptic function diverges at $u=1$, so that again the region around the upper integration limit will give the dominant contribution, and the elliptic function may be approximated by its asymptotic value \cite{Abramowitz}. This gives $v^{-1/2}\K\left(u/v^2\right)\approx-\ln\left(\delta/64\right)/4$. With these approximations, the integrals in Eq. (\ref{A2}) can be solved analytically, yielding Eq. (\ref{3.11}), with
\begin{align}
C\left(Y\right)=\frac{256\pi}{3}\frac{Y^2}{\Ein Y+6\ln2\left(\exp Y-1\right)}.\label{A3}
\end{align}
Here, $\Ein$ denotes the modified exponential integral \cite{Schelkunoff,Abramowitz}.

\section{TBR rate for arbitrary ratio $\bm{U_p/U_{ion}}$}\label{secC}
The energy distribution function of the free electrons in the presence of an RF field may be approximated by the shifted Boltzmann distribution
\begin{align}
f(U)=\frac{2\sqrt{U-U_p}}{\sqrt{\pi}(k_BT_e)^{3/2}}\exp\left(-\frac{U-U_p}{k_BT_e}\right)\Theta\left(U-U_p\right),\nonumber
\end{align}
where $\Theta$ denotes the Heaviside step function and the shift $U_p$ accounts for the quiver energy of the electrons. Substituting in Eq. (\ref{1.6}) this distribution function, the cross section $S_i(U)$ given in Ref. \cite{Hinnov}, and the rms velocity $v=\sqrt{2U/m}$ corresponding to energy $U$, and changing the integration variable to the thermal energy $U_{th}=U-U_p$, gives
\begin{align}R_i=&\frac{n_in_eme^4}{2\epsilon_0^2(2\pi mk_BT_e)^{3/2}}\int_{\left|U_i\right|}^\infty\left(\frac{1}{\left|U_i\right|+U_p}-\frac{1}{U_{th}+U_p}\right)\cdot\nonumber\\
&\cdot\sqrt{\frac{U_{th}}{U_{th}+U_p}}\exp\left(-\frac{U_{th}}{k_BT_e}\right)dU_{th}.\label{C2}
\end{align}
For field strengths $>1$ kV/m and typical UCP temperatures, $\left|U_i\right|>U_{ion}\gg k_BT_e$, so that the exponent in Eq. (\ref{C2}) falls off rapidly compared to the rate of variation of the pre-exponential factor; furthermore the integrand is only significant close to the lower integration limit. The pre-exponential factor may therefore be approximated by the first term of its Taylor-expansion around $U_{th}=\left|U_i\right|$. Performing the integration with this approximation, substituting for $n_i$ the equilibrium value from the Saha equation \cite{Mitchner}, and summing as before the result over all energy levels below $-U_{ion}$ by means of the rule $R=\sum R_i\approx\int R_iD(U_i)dU_i$ with $D(U_i)$ the density of states, gives the total TBR rate
\begin{align}
R&\approx\frac{\pi^2}{7g}\sqrt{\frac{2}{m}}\left(\frac{e^2}{4\pi\epsilon_0}\right)^5\frac{n_e^3}{U_{ion}^{7/2}k_BT_e}\:G\!\left(\frac{U_p}{U_{ion}}\right);\label{C3}\\
G(x)&\equiv \frac{7}{2x^3}\left(\frac{15+20x+3x^2}{3\left(x+1\right)^{3/2}}-\frac{5\arcsinh\sqrt{x}}{\sqrt{x}}\right).\label{C4}
\end{align}
The relative error in the approximation for the function $G(x)$ given in Eq. (\ref{1.10}) is less than $6\%$ for any value of $x$.

\section{Decay rate of $\bm{\nu_p}$ for UCP}\label{secB}
We use the following theorem \cite{Boyd}:\\
\emph{The coefficients of the Fourier series $\sum a_n\sin\left(n\Omega t\right)$ of a $2\pi/\Omega$-periodic function $y(t)$, which is infinitely many times differentiable, decay asymptotically as $a_n\propto \exp\left(-\Omega\tau n\right)$. The constant $\tau$ equals $\min\left|\im t_j\right|$, where $t_j$ denote the singularities of the function $y(t)$ in the complex $t$-plane.}\\

Writing $\omega=(\omega/\Omega)\cdot\Omega$ in Eq. (\ref{3.7}) shows that $X$ is essentially the $\omega/\Omega$-th Fourier coefficient of the function $x(\epsilon,t)$, so that according to the theorem the integrand in the collision frequency Eq. (\ref{3.10}) is proportional to
\begin{align}
\left|X\right|^2\propto\exp\left(-2\omega\tau\right);\hspace{1cm}\tau=\min\left|\im t_j\right|\label{B1}
\end{align}
for large $\omega$. This expression is easily checked for the model potential: the elliptic function in the trajectories Eq. (\ref{3.3}) has singularities along the lines $\im t=\pm \omega_1^{-1}\sqrt{2/v}\K\left(1-u/v^2\right)\equiv\pm\tau$ in the complex $t$-plane \cite{Lawden}. Substitution in Eq. (\ref{B1}) yields the behavior of $\left|X\right|^2$ for large $\omega$, which coincides precisely with what is found in Appendix \ref{secA}, Eq. (\ref{A1}) by explicit calculation.\\

\begin{figure}
\includegraphics[width=\columnwidth]{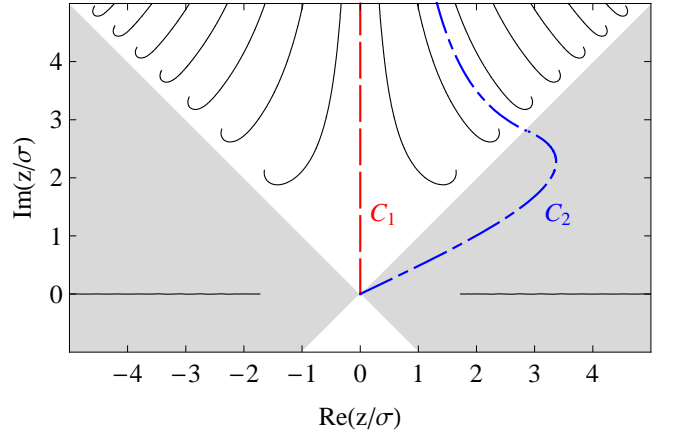}
\caption{\label{fig2} (Color online) Branch cuts (black solid lines) of the integrand of Eq. (\ref{B2}) in the complex $z$-plane, using $u=1$. In the shaded sectors $\left|\arg z\right|<\pi/4$ and $\left|\pi-\arg z\right|<\pi/4$, the error function behaves as $\erf\left(z/\sigma\right)\rightarrow 1$ as $\left|z/\sigma\right|\rightarrow\infty$. Two possible contours from the origin to infinity are shown.}
\end{figure}

Applying Eq. (\ref{B1}) to the actual UCP potential requires explicit expressions for the trajectories $x(\epsilon,t)$, however these are not known. Instead, the inverse function $t(\epsilon,x)$ may be obtained by integration of the equation of motion $md^2x/dt^2=-dU(x)/dx$, yielding
\begin{align}
t(\epsilon,x)=\sqrt{\frac{m}{2}}\int_0^x\frac{dz}{\sqrt{\epsilon-U(z)}}\label{B2}.
\end{align}
Here, the initial conditions $x=0$ and $dx/dt=\sqrt{2\epsilon/m}$ at $t=0$ have been assumed, and $U(z)$ denotes the UCP potential Eq. (\ref{3.1}) with $r=z$. Equation (\ref{B1}) requires knowledge of the singularities $t_j$ of the functions $x(\epsilon,t)$, which may be categorized as either poles, logarithmic branch points or algebraic branch points. (More pathological singularities such as $\exp\left(1/z\right)$ at $z=0$ are not considered here.) An algebraic branch point in $x(\epsilon,t)$ corresponds to a critical point in the inverse function $t(\epsilon,x)$, at which $dt/dx=0$. Differentiating Eq. (\ref{B2}) with respect to $x$, it follows that $U(z)$ must diverge at such a point if the derivative $dt/dx$ is to vanish. But the UCP potential Eq. (\ref{3.1}) is an entire function, so that this does not occur for any finite complex $z$, hence $x(\epsilon,t)$ does not have any algebraic branch points.\\

\begin{figure}
\includegraphics[width=0.8\columnwidth]{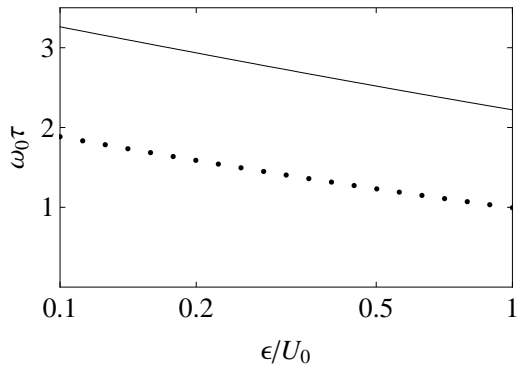}
\caption{\label{fig3} Decay constant $\tau$ in Eq. (\ref{3.11}) as a function of particle energy. Solid line: analytical result for the model potential Eq. (\ref{3.1}) assuming $\omega_1=\omega_0$ and $U_1=U_0$; dots: numerical result for the UCP potential (\ref{3.2})}.
\end{figure}

Considering next poles and logarithmic branch points in $x(\epsilon,t)$, at such points the position diverges while the complex time has some finite value. In terms of the inverse function Eq. (\ref{B2}) then, there exist contours $C_j$ in the complex $x$-plane from the origin to infinity such that $t(\epsilon,x)\rightarrow t_j$ with $\left|t_j\right|<\infty$ as $x\rightarrow\infty$ along $C_j$. In view of Eq. (\ref{B1}) we are interested in the contour that yields the time $t_j$ with the smallest imaginary part. A complication in finding this contour is the presence of the square root in Eq. (\ref{B2}), because of which the integrand has branch cuts in the complex $z$-plane. Adopting the standard choice of letting the branch cuts coincide with the points at which the argument of the root is real and negative, these cuts start at the zeros of the function $\epsilon-U(z)$ and extend to $\pm i\infty$ without crossing. Fig. \ref{fig2} shows the resulting branch cut structure for the case $\epsilon=U_0/2$; the integrand in the lower half-plane is the complex conjugate of that in the upper half-plane. Also drawn are two possible contours from the origin to infinity. Now, the potential $U(z)$ in Eq. (\ref{B2}) contains the error function $\erf\left(z/\sigma\right)$, which has the property \cite{Abramowitz} that its value is close to unity for $\left|z/\sigma\right|\gtrsim 1$ in the shaded sectors in Fig. \ref{fig2}, while its amplitude grows superexponentially as $z\rightarrow\infty$ in the non-shaded sectors. Therefore the integrand in Eq. (\ref{B2}) will be essentially constant along parts of contours that cross the shaded sector, such as $C_2$, so that a large contribution to the integral is accumulated along these parts. Hence we may expect that the contour yielding the smallest possible value of $t_j$ is the contour that avoids the shaded sectors altogether, that is, the contour $C_1$ along the imaginary axis. With this conjecture, we calculate $\tau$ in Eq. (\ref{B1}) by integrating Eq. (\ref{B2}) along $C_1$ for several values of the particle energy $\epsilon$. The result is shown in Fig. \ref{fig3}, together with the analogous result for the model potential. As is clear from the figure, for any particle energy $\tau$ for the UCP potential is approximately one unit $\omega_0^{-1}$ less than that for the model potential. Hence, asymptotically for large $\omega$, the quantity $\left|X\right|^2$ in Eq. (\ref{3.10}) will contain an extra factor $\exp\left(2\omega/\omega_0\right)$ as compared to the case of the model potential, independent of $\epsilon$. The resulting rate of decrease of the collision frequency is given in Eq. (\ref{3.12}).

\end{document}